# Harmonica: A Framework for Semi-automated Design and Implementation of Blockchain Applications

Nicolas Six, nicolas.six@univ-paris1.fr,
Nicolas Herbaut, nicolas.herbaut@univ-paris1.fr,
Camille Salinesi, camille.salinesi@univ-paris1.fr

## Abstract

Designing blockchain-based applications is a tedious task. Compared to traditional software engineering, software architects cannot rely on previous experiences or proven practices, often formalized as software patterns. Also, the selection of an adequate blockchain technology is difficult without deep knowledge of the technology. This paper introduces Harmonica, a framework for the design and implementation of a blockchain-based application. This framework is divided in three parts: a decision-making engine to recommend a blockchain technology and blockchain-based software patterns relying on requirements, a configurator to generate code stubs and configuration files, and a knowledge base to support those tools.

## Introduction

A blockchain is a ledger containing transactions embedded in linked blocks and shared by a network of nodes. In this network, an algorithm run by every node manages the inclusion of new transactions to the blockchain, namely the consensus algorithm. A transaction is an operation that changes the state of the blockchain. They are often used for two types of operation: exchanging cryptocurrencies between users and interacting with smart-contracts, which are programs stored on-chain that can also perform operations to change the state of the blockchain.

Through their special operating models, blockchain technologies have many unique characteristics. First, a blockchain network is disintermediated: no one is fully responsible for the management of the network. Second, using blockchain provides data security and immutability. The addition of blocks is ruled by a consensus algorithm and alteration of data is impossible, where traditional data storage technologies can be modified by authorized parties. Third, it is possible to retrace the complete history of state changes of a blockchain, enabling full traceability. However, the technology suffers from several drawbacks. Due to its design, blockchain often suffers from scalability and performance issues. Where traditional databases can meet hundreds of thousands of transactions per second, most of the public blockchains cannot reach even a fraction of it. Also, every data inside is publicly available, which can be an issue when data exchanges must be kept secret between users. The immutability of data is also a problem when using smart-contracts: upgrading an already deployed smart-contract is often impossible by design and specific techniques must be used.

In recent years, many new blockchain technologies have been designed to tackle those issues, but there is a constant balance between strengths and liabilities. For example, some blockchains (private blockchains) only allow a specific set of known nodes to join the network, add new blocks, and form a consortium, where others (public blockchains) let any user join the network and start creating blocks through slow albeit robust consensus algorithms, such as Proof-of-Work. Some blockchains also integrate data deletion features (automatic pruning), but this is in opposition with the immutability of blockchains. As the technology emerges and diverges from others, blockchain integration in new or existing software and systems is still a challenge. Most practitioners do not have enough expertise in the field to decide on which blockchain solution to use for a specific context. They must also be aware of blockchain technical specificities that differ from conventional technologies (high latency, data access rules, impossibility to query data from outside the blockchain, …). Where using architectural or design patterns to build software is a widespread practice in the software engineering field (Devedzic 2002), it is not in blockchain-based architectures, where only a few have been proposed and might lack extensive testing. Finally, it is also difficult to bootstrap a blockchain system from scratch if the practitioner chose to use a private blockchain. An adequate initial configuration is paramount to satisfy many system requirements (for example, performance) and is hard to update on a running blockchain network.

To address those issues, engineers designed models and tools to assist the practitioner in the choice of blockchain technology (Belotti et al. 2019). Engineers also created new patterns to support the design of parts of the application (Xu et al. 2018), and designed software to generate code stubs of smart contracts (Frantz and Nowostawski 2016). However, there is still no holistic framework yet to assist practitioners from the design to the development of blockchain applications. This paper proposes Harmonica, an end-to-end framework to fill the gap, through a suite of tools and a knowledge base. The next section introduces in detail the framework and its content, then a conclusion is given, and future work using Harmonica.

**Framework Presentation**

The proposed framework (Figure 1) is divided into two main parts, respectively the tooling suite and the knowledge base. The tooling suite is composed of two tools: (1) BLADE, a decision-making tool for the selection of a blockchain and blockchain-based patterns in a given context, and (2) BANCO, a configurator designed to assemble the major parts of a blockchain application using the software product line principles. Provided tools can be called independently from each other, an architect can obtain recommendations, code stubs and configuration files, or both. The framework's tools leverage a knowledge base, which contains information about a set of blockchain technologies and blockchain-based patterns, as well as core assets (for example, configurable code stubs) to build the software at the end.

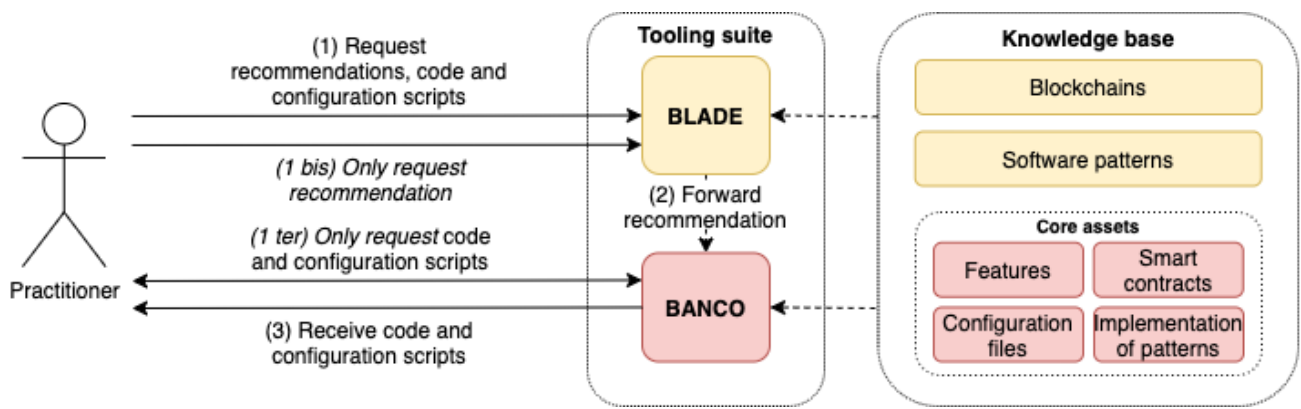

*Figure 1. Framework overview (red artifacts are currently in development and yellow artifacts have already been implemented but further developments are planned to improve them)*

**Knowledge base**

To perform, the tooling suite relies on a knowledge base, divided into three subsets:

- Blockchains: this subset contains organized knowledge about existing blockchain technologies. Each blockchain is described by attributes that allow practitioners to differentiate them from each other and gives detail on their capabilities.

- Software patterns: this subset contains the blockchain-based patterns (for decision-making). A template will be implemented from those patterns and stored with other core assets for code generation.

- Core assets: these are reusable elements to construct the blockchain application at the end. It contains smart-contracts, code features, blockchain configuration files, and implementation of patterns.

Building an efficient and useful knowledge base requires finding a suitable format to store collected data. Blockchain patterns are organized into an ontology that describes the different concepts (blockchain and patterns), and the relations between them. Such an approach helps to make recommendations, as powerful reasoning between concepts can be performed. Eventually, blockchain data will also be included in a dedicated ontology, allowing inferences between the two ontologies. To fill the knowledge base, there was consideration of multiple approaches. For the software pattern ontology, a systematic literature review identified existing blockchain-based patterns, and a taxonomy built from acquired patterns. Another envisioned approach is collaborative editing, to acquire knowledge from contributors that have an interest in the result. Where such methods are sufficient for the construction of a pattern knowledge base, they might not be efficient enough to build an accurate blockchain knowledge base, where blockchains are frequently updated, leading to changes in their attributes. Another considered approach to tackle this issue is the use of automatic methods such as scrapping or natural language processing (NLP) to collect knowledge of relevant documents (for example, whitepapers, academic literature, …). The build of this first version of this knowledge base supports the first iterations of BLADE and is published on GitHub.

**BLADE**

Relying on the knowledge base, BLADE is a tool capable of suggesting the most suitable blockchain and blockchain-based patterns to use for a given context. So-called context is an aggregation of different inputs: user requirements, models, and the company's assets (for example, infrastructure definition).

The process to generate recommendations is the following. First, the user must specify the blockchain attributes desired or required for the decision-making. The user can select a label (from *Indifferent* to *Extremely Desirable*) to express its level of preference towards an attribute. The user can also specify if an attribute is *Required*; if so, a blockchain that does not meet this requirement is automatically disqualified. BLADE dynamically generates a dependency model when a user selects requirements to prevent the user from selecting two requirements that conflict between each other. For example, it helps to balance the different strengths and liabilities of blockchain, such as immutability versus modifiability, or decentralization versus access control. We implemented the first version of BLADE for decision-making between five blockchain technologies (Six, Herbaut, and Salinesi 2020) described by 14 non-functional requirements as attributes, and a multi-criteria decision-making algorithm named TOPSIS (Lai and Hwang 1994).

**BANCO**

BANCO is the third artifact constituting this framework, to generate code and scripts from the recommendations, requirements, and user selection of features. BANCO leverages a variability model for the selection of many features that will compose the final product. A user can access further assistance for selection from the recommendations produced by BLADE, but also use both tools independently. Using the assistance, many features will be preselected such as the blockchain solution to use and recommended patterns. Following that, BANCO will use a library of core assets (code templates, configuration files, …) to generate parts of the new blockchain application such as smart-contracts or off-chain components for blockchain interaction. Each of those artifacts will contain off-the-shelf features, also generated from existing core assets. Using this approach also allows the generation of scripts to bootstrap a private blockchain with adequate configuration on multiple machines automatically. This work is still in its initial design and will be implemented in a later stage.

**Conclusion**

This framework aims at facilitating the work of software and systems engineers for the design and the implementation of blockchain applications, by proposing a collection of tools to obtain recommendations and artifacts to build the application over and set up the system with ease. Future works will consist in improving the existing artifacts and developing the others. First, by enhancing BLADE with the support of pattern decision-making. Using BLADE, architects will be able to get precise recommendations on the blockchain and related patterns to use, based on a knowledge base and a systematic process. Then, the plan is for the addition of more alternatives into the knowledge base, as well as the core assets and the architectural patterns to make more accurate decisions using BLADE and support the generation of code with BANCO. For example, an ongoing systematic literature study to collect patterns from existing solutions proposed by researchers will serve to add new patterns into the knowledge base. Finally, with the implementation of BANCO, Harmonica will take the recommendations of BLADE and the requirements of the user to generate code stubs that can be used off-the-shelf or customized to develop a blockchain application. We expect to trial the framework with blockchain experts and software or system architects to validate its correctness and utility when applied, notably through case studies on different domains.